\documentclass[twocolumn,showpacs,preprintnumbers,amsmath,amssymb]{revtex4}

\usepackage{graphicx}
\usepackage{dcolumn}
\usepackage{bm}

\begin{document}

\preprint{APS/123-QED}

\title{Measurement of the 20 and 90 keV resonances in the ${}^{18}{\rm O}(p,\alpha){}^{15}$N
reaction via THM}

\author{M. La Cognata$^1$}
\author{C. Spitaleri$^1$}
\email{Spitaleri@lns.infn.it}
\author{A.M. Mukhamedzhanov$^2$}
\author{B. Irgaziev$^3$}
\author{R.E. Tribble$^2$}
\author{A. Banu$^2$}
\author{S. Cherubini$^1$}
\author{A. Coc$^4$}
\author{V. Crucill\`a$^1$}
\author{V.Z. Goldberg$^2$}
\author{M. Gulino$^1$}
\author{G.G. Kiss$^5$}
\author{L. Lamia$^1$}
\author{J. Mrazek$^6$}
\author{R.G. Pizzone$^1$}
\author{S.M.R. Puglia$^1$}
\author{G.G. Rapisarda$^1$}
\author{S. Romano$^1$}
\author{M.L. Sergi$^1$}
\author{G. Tabacaru$^2$}
\author{L. Trache$^2$}
\author{W. Trzaska$^7$}
\author{A. Tumino$^1$}
\affiliation{
$^1$INFN Laboratori Nazionali del Sud \& DMFCI Universit\`a di Catania Catania Italy\\
$^2$Cyclotron Institute Texas A\&M University College Station TX USA\\
$^3$GIK Institute of Engineering Sciences and Technology Topi District Swabi NWFP Pakistan\\
$^4$CSNSM CNRS/IN2P3 Universit\`e Paris Sud Orsay France\\
$^5$ATOMKI Debrecen Hungary\\
$^6$Nuclear Physics Institute of ASCR Rez near Prague Czech Republic\\
$^7$Physics Department University of Jyvaskyla Finland
}

\date{\today}

\begin{abstract}
The $^{18}{\rm O}(p,\alpha)^{15}{\rm N}$ reaction is of primary
importance in several astrophysical scenarios, including fluorine
nucleosynthesis inside AGB stars as well as oxygen and nitrogen
isotopic ratios in meteorite grains. Thus the indirect measurement
of the low energy region of the $^{18}{\rm O}(p,\alpha)^{15}{\rm
N}$ reaction has been performed to reduce the nuclear uncertainty
on theoretical predictions. In particular the strength of the 20 and 90
keV resonances has been deduced and the change in the reaction rate
evaluated.
\end{abstract}

\maketitle

Understanding fluorine production would make predictions on
asymptotic giant branch (AGB) star nucleosynthesis more
accurate. The AGB stage represents the final nuclosynthetic phase
in low and intermediate mass stars. AGB stars play an extremely 
important role in astrophysics because heavy elements along the 
stability valley are produced in their interiors through slow
neutron captures (s-process). $^{19}$F can be produced during the thermal
pulse that is ignited in the $^4$He-rich intershell region,
following the ingestion of the $^{13}$C pocket. The subsequent
third dredge-up (TDU) episode mixes the products of shell flash
He-burning, including fluorine, and $s$-process nuclei to the outer layers. 
Therefore $^{19}$F abundance constitutes a key parameter to constrain
AGB star models \cite{LUG04}. Anyway, if the theoretical
abundances are compared to the observed ones, a remarkable discrepancy
shows up because the largest $^{19}$F abundances cannot be matched
for the typical $^{12}{\rm C}/^{16}{\rm O}$ ratios \cite{LUG04}. 
A possible way to explain $^{19}$F abundance and several related 
observables (such as isotopic ratios in meteorite grains \cite{NOL03}) 
can be linked to $^{18}$O as the $^{18}$O$(p,\alpha){}^{15}$N reaction 
represents the main $^{15}$N production channel, both in the intershell 
region and at the bottom of the convective envelope \cite{NOL03,LUG04}. 
During the thermal pulse $^{15}$N is burnt to $^{19}$F via the
$^{15}$N($\alpha$,$\gamma$)$^{19}$F reaction. Thus a larger
$^{18}$O$(p,\alpha){}^{15}$N reaction rate would lead to an
increase of the $^{19}$F supply as well as to an enrichement of
$^{15}$N in the stellar surface, which would play a key role
to explain the long-standing problem of the $^{14}{\rm N}/^{15}{\rm N}$ 
ratio in meteorite grains (\cite{NOL03} and references therein).

The $^{18}$O$(p,\alpha){}^{15}$N reaction has been the subject of
several experimental investigations \cite{MAK78,LOR79} and many
features are known from spectroscopic studies
\cite{YAG62,CHA86,WIE80,SCH70}. Nevertheless the reaction rate for
the process has a considerable uncertainty \cite{ANG99}. In the
$0-1000$ keV energy range, where 9 resonances occur, the reaction
rate is essentially determined by the 20 keV, 144 keV and the 656
keV resonances \cite{ANG99}. 
With regard to the 20 keV resonance, its strength is known only
from spectroscopic measurements performed through the transfer reaction
$^{18}{\rm O}(^{3}{\rm He},d)^{19}{\rm F}$ \cite{CHA86}
and the direct capture reaction $^{18}{\rm O}(p,\gamma)^{19}{\rm F}$
\cite{WIE80}. Such estimates, which are based on the
deduced spectroscopic factors, are strongly model dependent 
(being connected to the adopted optical model potentials) and
affected by large and not-well-defined uncertainties.
An additional important source of uncertainty on the reaction rate
is connected with the determination of the resonance energy \cite{CHA86}.
Furthermore the spin and parity of the 8.084 MeV level in
$^{19}$F (corresponding to a 90 keV resonance in the
$^{18}$O$(p,\alpha){}^{15}$N cross section) has not been
established. The uncertainties on nuclear physics inputs have made
astrophysical predictions far from conclusive \cite{NOL03}.

In order to reduce the nuclear uncertainties affecting the
reaction rate estimate we have performed an experimental study of
the $^{18}$O$(p,\alpha){}^{15}$N reaction by means of the Trojan
horse method (THM), which is an indirect
technique to measure the relative energy-dependence of a charged-particle reaction cross
section at energies well below the Coulomb barrier
(\cite{LAC07,SPI99} and references therein). The cross section of
the relevant $A+x \to c+C$ reaction is deduced from a suitable
$A+a(x \oplus s) \to c+C+s$ process, performed in quasi-free (QF)
kinematics. The beam energy is chosen larger than the Coulomb
barrier for the interacting nuclei, so the break-up of nucleus $a$
(the so-called $Trojan$-$horse$) takes place inside the $A$
nuclear field. Therefore, the cross section of the $A+x \to c+C$
reaction is not suppressed by the Coulomb interaction of the
target-projectile system. 

In a previous investigation, carried out at the Cyclotron Institute, 
Texas A\&M University, Texas (USA) \cite{LAC08}, the $^{18}$O$(p,\alpha){}^{15}$N
reaction has been measured via the THM through the $^2$H(${}^{18}{\rm
O},\alpha{}^{15}$N)$n$ process in the $0-1000$~keV $^{18}{\rm O} - p$ relative
energy range. For the first time the energy
region below 70 keV had been investigated. In the present work we focus on a new
study of the $^{18}$O$(p,\alpha){}^{15}$N reaction by means of the
same THM process. The aim is to span the $^{18}{\rm O} - p$
relative energy region below 250 keV with an improved energy
resolution, in order to deduce resonance parameters and J$^{\pi}$ values of the 8.014
and 8.084 MeV $^{19}$F levels.
The experiment was performed at Laboratori Nazionali del Sud,
Catania (Italy). The SMP Tandem Van de Graaff accelerator provided the 54 MeV
$^{18}$O beam impinging onto thin self-supported deuterated polyethylene (CD$_2$)
targets. The detection setup consisted of a telescope (A), devoted to
$^{15}$N detection, made up of an ionization chamber and a silicon
position sensitive detector (PSD A) on one side with respect to
the beam direction, and three additional silicon PSD's (B, C and D) on the
opposite side. Angular conditions were selected in order to maximize the expected QF
contribution. Channel selection has been accomplished by gating on
the $\Delta E - E$ 2D spectra to select the nitrogen locus.


Compelling evidence for the occurrence of the QF mechanism is 
given by the shape of measured momentum-distribution,
if it follows the shape of the deuteron wave function.
In the analysis, the theoretical distribution, given by the square of the Hulth\'en wave function
in momentum space in the plane wave approximation \cite{LAC07}, 
is superimposed onto the experimental one. The
good agreement demonstrates that the QF mechanism is present and
dominant in the $|p_3|<50$ MeV/c neutron momentum range. 
Thus, in the following analysis
only the phase space region for which the $|p_3|<50$ MeV/c condition
is satisfied is taken into account.

\begin{figure}[!t]
\includegraphics[scale=0.45]{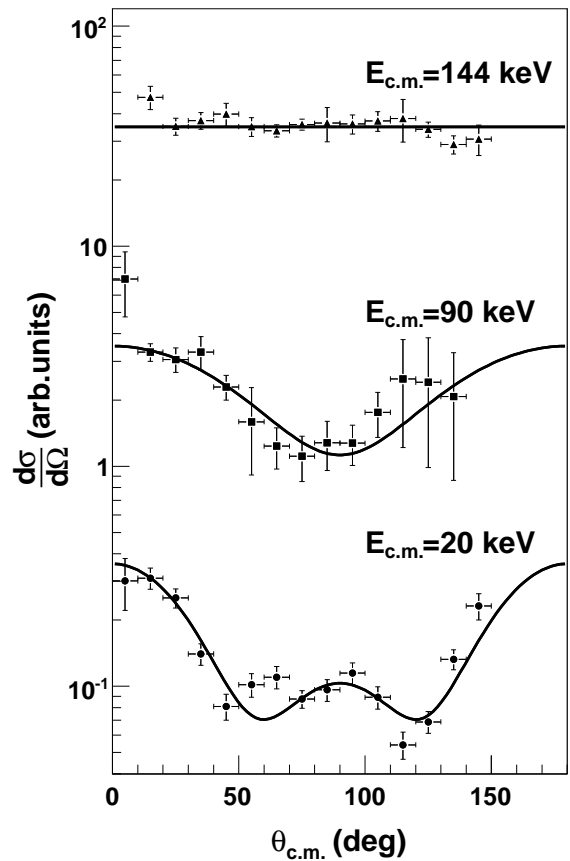}
\caption{\label{angdistr} Experimental angular distributions for the
$^{18}{\rm O}(p,\alpha){}^{15}{\rm N}$ reaction for the three resonances
in the $0-250$ keV energy range. The full lines come from the fitting
with the curve of Eq. \ref{angfit}.}
\end{figure}

Angular distributions for the $^{18}$O$(p,\alpha){}^{15}$N reaction were
extracted as discussed in \cite{LAC07}.
They were used to perform the necessary validity test on the
deduced cross section and to infer the spin-parity for the states.
The $\alpha$ emission angles in the c.m. system ($\theta_{c.m.}$),
covered in the present experiment,
were about $\theta_{c.m.}$ = 0$^{\circ}$ - 60$^{\circ}$,
$\theta_{c.m.}$ = 40$^{\circ}$ - 110$^{\circ}$ and $\theta_{c.m.}$
= 90$^{\circ}$ - 150$^{\circ}$ for the A-B, A-C and the
A-D detector coincidences respectively. The presence of an overlap region
allowed for relative normalization between the
cross sections deduced from each couple. Angular distributions
were extracted for several energies, focusing in particular on the
20, 90 and 143.5 keV resonances. The results are displayed in Fig. \ref{angdistr}
where the experimental data are given as filled circles (20 keV), squares
(90 keV) and triangles (143.5 keV). Errors on the half-off-energy-shell 
(HOES) \cite{LAC07} cross section $\frac{d\sigma}{d\Omega}$ 
account for statistics and for the deconvolution of the single
resonance contributions. The error on $\theta_{c.m.}$ represents the width of
each bin (chosen in order to have reasonable statistical precision).

From Fig. \ref{angdistr} it turns out that
the ${\rm J}^{\pi}=\frac{1}{2}^+$  assignment for the 143.5 keV resonance
is confirmed, the angular distribution for that level being isotropic
(see Fig. \ref{angdistr}).
This result represents a cross check of the method, since we are able to
reproduce the angular distribution for a well known resonance. Therefore
we extracted the angular distribution for both the 90 keV
and 20 keV resonances. A fit of the experimental data was performed by:
\begin{eqnarray}
\frac{d\sigma}{d\Omega_{c.m.}}=\frac{d\sigma}{d\Omega_{c.m.}}(90^{\circ})
\left(1+A_2(E)\cos^2\theta \right. \nonumber \\
\left. +A_4(E)\cos^4\theta+...+A_{2L}(E)\cos^{2L}\theta\right)
\label{angfit}
\end{eqnarray}
where $L$ is the smallest among the spin J of the formed compound nucleus
and the angular momentum $l$ quantum numbers in the
entrance and the exit channels \cite{EVA69}.
The best fit for the 90 keV resonance is achieved for $L=1$ ($\tilde{\chi}^2=0.67$), thus
we infer a spin for the 8.084 MeV excited state of $^{19}$F of $\frac{3}{2}$.
We consider that this $L$ value corresponds to the angular momentum in the
exit channel, because of considerations on the proton width of this level, 
and the parity to be positive (the $^{15}$N ground state being  $\frac{1}{2}^-$).
For the 20 keV resonance, the best fit is obtained for
$L=2$ ($\tilde{\chi}^2=3.11$), supporting the J$^{\pi}=\frac{5}{2}^+$ spin-parity
assignment for the $^{19}$F excited state at 8.014 MeV (see \cite{CHA86,SCH70}).
In the extraction of the angular distributions the $^{18}{\rm O} -p$ relative
energy is kept fixed thus preventing possible bias caused by the HOES nature of the deduced cross
section. The HOES effect might be expressed by a constant renormalization
factor, and therefore the angular distributions are given in
arbitrary units in Fig. \ref{angdistr}.

\begin{figure}[!t]
\includegraphics[scale=0.45]{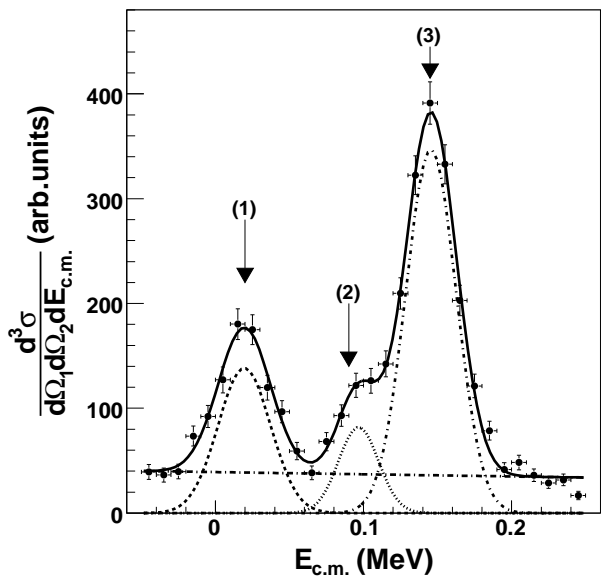}
\caption{\label{sezurtotot} Cross section of the TH reaction (full circles). (1), (2) and (3)
refer to the 8.014, 8.084 and 8.138 MeV $^{19}$F levels. The full
line represent the result of a fit including three Gaussian curves and a 1st order polynomial.}
\end{figure}

The extracted HOES differential cross section
has been integrated in the whole angular range. It was assumed that in the
region where no experimental angular distributions are available, their trend
is given by the fit of Eq. \ref{angfit}. The
resulting $^{2}{\rm H}(^{18}{\rm O},\alpha^{15}{\rm N})n$ reaction
cross section is shown in Fig. \ref{sezurtotot}  (full circles). The experimental
energy resolution turned out to be about 40 keV (FWHM), in agreement with the
value predicted by Monte Carlo simulations. Horizontal
error bars represent the integration bin while the vertical ones arise
from statistical uncertainty and angular distribution
integration. Indeed, because of the limited energy resolution, resonances were not well
resolved thus each contribution to the reaction yield had to be disentangled
when extracting angular distributions. The solid line in the figure is the sum
of three Gaussian functions to fit the resonant behavior and a straight line
to account for the non-resonant contribution to the cross section. This
fit was performed with the sole aim of extracting the resonance energies: ${\rm E}_{R_1} = 19.5 \pm 1.1$ keV,
${\rm E}_{R_2} = 96.6 \pm 2.2$ keV and ${\rm E}_{R_3} = 145.5 \pm 0.6$ keV (in fair
agreement with the ones reported in the literature \cite{ANG99}) and
to deduce the peak values of each resonance: $N_1 = 138 \pm 8$, $N_2 = 82 \pm 9$
and $N_3 = 347 \pm 8$ (arbitrary units). The peak values were used to derive 
the resonance strengths $\omega\gamma$ that are the relevant parameters for astrophysical application
in the case of narrow resonances \cite{ANG99}. 
The THM cross section for the $A+a(x \oplus s) \to c+C+s$ reaction
proceeding through a resonance $F_{i}$ in the subsystem $F =A + x = C + c$ 
is \cite{LAC07,muk2007}
\begin{equation}
\frac{d^2\sigma}{dE_{Cc}\,d\Omega_s} \propto 
\frac{\Gamma_{(Cc)_{i}}(E)\,|M_{i}(E)|^{2}
}{(E - E_{R_{i}})^{2} + \Gamma_{i}^{2}(E)/4} .
\label{d3sigma}
\end{equation}
Here, $M_{i}(E)$ is the direct transfer reaction amplitude for the 
binary reaction $A + a \to F_{i} + s$  populating the resonant state 
$F_{i}$ with the resonance energy $E_{R_{i}}$,  $\,E$ is the $A-x$ relative kinetic energy 
related to $E_{Cc}$ by the energy conservation law, $\,\Gamma_{(Cc)_{i}}(E)$ is the partial 
resonance width for the decay $ F_{i} \to C + c$ and $\Gamma_{i}$ is the total 
resonance width of $F_{i}$. The appearance of the transfer reaction amplitude 
$M_{i}(E)$ instead of the entry channel partial resonance width $\Gamma_{(Ax)_{i}}(E)$ 
is the main difference between the THM cross section and the cross section for the 
resonant binary sub-reaction $A + x \to C + c$ \cite{LAC07,muk2007}.
On the other hand, the resonance parameters deduced by means of the THM are not affected
by the electron screening, distorting the direct S-factor at the astrophysically
relevant energies \cite{ASS87}. 

\begin{figure}[!t]
\includegraphics[scale=0.44]{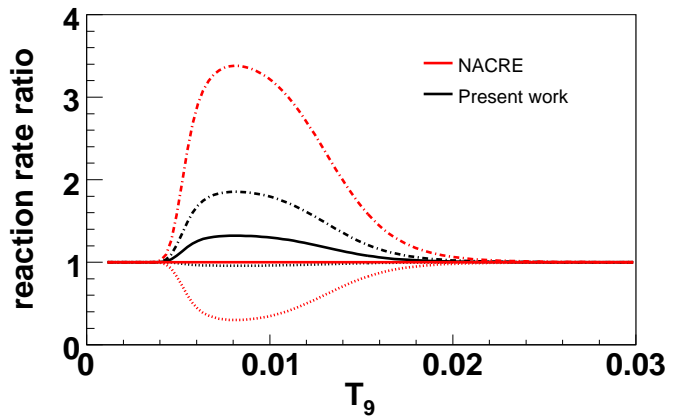}
\caption{\label{rateratio} Ratio of the THM reaction rate (full black line) 
to the NACRE one \cite{ANG99}. For comparison, the NACRE rate is shown as a red line. 
The dot-dashed and dotted lines show the upper and lower limits allowed by
experimental uncertainties.}
\end{figure}

The peak THM cross section taken at the $E_{R_{i}}$  resonance
energy for the $(p,\alpha)$ reaction $A + x \to C + c$ is given by
\begin{equation}
N_i =  4\,\frac{\Gamma_{\alpha_{i}}(E_{R_i})\,M_{i}^2(E_{R_i})}{\Gamma_{i}^{2}\,
(E_{R_{i}})}, 
\label{maxval}
\end{equation}
where $\Gamma_{Cc_{i}}(E) \equiv \Gamma_{\alpha_{i}}(E)$. 
In this work we did not extract the absolute value of the cross section. The proton and 
alpha partial widths for the third resonance at 143.5 keV are well known \cite{ANG99}, thus we can determine 
the alpha partial width and the strength for the first and second resonances from the ratio of the peak 
values of the THM cross sections:
\begin{equation}
(\omega\gamma)_{i}=\frac{\omega_{i}}{\omega_{3}}
\frac{\Gamma_{p_{i}}(E_{R_{i}})}{|M_{i}(E_{R_{i}})|^2}
\frac{|M_{3}(E_{R_{3}})|^2}{\Gamma_{p_{3}}(E_{R_{3}})}
\frac{N_i}{N_{3}}(\omega\gamma)_{3}, \quad  i=1,2.
\label{ogthm}
\end{equation}
Here, $(\omega\gamma)_{i}=\omega_{i}\,\Gamma_{\alpha_{i}}(E_{R_i})\,\Gamma_{p_{i}}(E_{R_{i}})/\Gamma_{i}(E_{R_{i}})$ 
is the $i$-th resonance strength, $\omega_{i}= {\hat J}_{i}/({\hat J}_{A}\,{\hat J}_{x})$ is 
the statistical factor, ${\hat J}= 2\,J + 1$, $\,J_{i}$ is the spin of the $i$-th resonance 
and $J_{A},\,J_{x}$ are the spins of $A$ and $x$, respectively. 
When determining $(\omega\gamma)_{i}$  the effect of energy resolution 
in our experiment has been taken into account. The electron screening gives a negligible contribution around 144 keV
(4\% maximum \cite{ASS87}), thus no systematic uncertainty is introduced by normalizing to the third resonance.
In the plane wave approximation 
$M_{i} \approx  \varphi_{a}(p_{sx})\,W_{Ax}({\rm {\bf p}}_{Ax})$, where $\varphi_{a}(p_{sx})$ is the Fourier transform 
of the $s$-wave radial $s-x$ bound wave function, $p_{sx}$ is the $s-x$ relative momentum, and 
$W_{Ax}= <I^{F^{*}}_{Ax}|V_{Ax}|{\rm {\bf p}}_{Ax}>$ is the form factor for the synthesis 
$A + x \to F_{i}$, $I^{F^{*}}_{Ax}$ is the overlap function of the bound state wave functions 
of $A,\,x$  and the resonant wave function of $F^{*}$, ${\rm {\bf p}}_{Ax}$ is the $A-x$ relative 
momentum. In  practical calculation we approximated $I^{F_{i}}_{Ax}$ by $S_{i}^{1/2}\,\varphi_{(Ax)_{i}}$, 
where $S_{i}$ is the spectroscopic factor for configuration $F_{i}=A + x$ and $\varphi_{(Ax)_{i}}$ is 
the single-particle bound-state type wave function describing the resonance state $F_{i}$. 
Since in Eq. \ref{ogthm} only the $\frac{\Gamma_{p_{i}}(E_{R_{i}})}{|M_{i}(E_{R_{i}})|^2}$
ratios show up, the dependence on spectroscopic factors and proton widths is completely removed and,
as a consequence, $(\omega\gamma)_{i}$ is connected to $(\omega\gamma)_{3}$ through the easily calculable
ratios of the single-particle widths to the form factors $W_{Ax}$.
If $(\omega\gamma)_3$ is taken from \cite{BEC95}, by means of Eq. \ref{ogthm} 
one gets $(\omega\gamma)_1=8.3^{+3.8}_{-2.6}\times 10^{-19}$ eV,
which is well within the upper and lower limits given by NACRE, $6^{+17}_{-5}\times 10^{-19}$ eV \cite{ANG99}.
While NACRE recommended value is based on varius kinds of estimates, the present result
is obtained from experimental data, thus the accuracy of the deduced resonance 
strength has been greatly enhanced.
The largest contribution to the error is due to the uncertainty on the resonance energy, 
while statistical and normalization errors sum up to about 9.5\%.
With a similar approach we have obtained
$(\omega\gamma)_2=(1.76 \pm 0.33)\times 10^{-7}$ eV (statistical and 
normalization errors $\sim$ 13\%) for the 90 keV resonance,
in good agreement with the strength given by NACRE, $(1.6 \pm 0.5)\times 10^{-7}$ eV \cite{ANG99}.
This gives us confidence in the theory used in the THM allowing for a cross check of the method.

By using the narrow resonance approximation \cite{ANG99}, which is fulfilled for the resonances under investigation,
the reaction rate has been deduced and compared with the one reported in NACRE
\cite{ANG99}. In Fig. \ref{rateratio} the ratio of the THM reaction rate to
the NACRE one for the $^{18}{\rm O}(p,\alpha){}^{15}{\rm N}$ is shown as a full black
line. The dot-dashed and dotted black lines represent the upper and lower limits respectively, allowed
by the experimental uncertainties. In the low temperature region (below
T$_9=0.03$, Fig. \ref{rateratio}a) the reaction rate can be about 35\% larger 
than the one given by NACRE (full red line), while the indetermination is greatly reduced with respect
to the NACRE one (dot-dashed and dotted red lines mark the upper and lower limits, Fig. \ref{rateratio}).
Those temperatures are typical of the bottom of the convective envelope, thus an increase of
this reaction rate might have important consequences on the cool bottom process
\cite{NOL03} and, in turn, on the surface abundances and isotopic ratios in AGB stars.
The 8.084 MeV excited state of $^{19}$F (corresponding to the 90 keV resonance) provides a negligible
contribution to the reaction rate in agreement with
previous estimate \cite{CHA86}.

In conclusion, in this paper we demonstrated, for the first time, the power 
of indirect THM which allowed to determine the strength of the low-lying 20-keV resonance 
in ${}^{19}{F}$, elusive for any direct technique. This resonance plays an important role in the 
determination of the reaction rates of the key astrophysical reaction
$^{18}{\rm O}(p,\alpha){}^{15}{\rm N}$. We also presented a new way of determination 
of the resonance strength by measuring the relative strengths of the known and 
unknown resonances and using the half-off-sell R-matrix determine the strength 
of the unknown resonance avoiding information about the spectroscopic factors.  
The parameters of the 20 keV resonance in the $^{18}{\rm O}(p,\alpha){}^{15}{\rm N}$ reaction
relevant for astrophysics together with those for the nearby resonances could
be extracted and importantly their uncertainty was strongly reduced (by a factor $\sim 8.5$), thanks to
the newly developed approach, which is based on experimental 
data in contrast to the NACRE one that relies on varius
kinds of estimate. Indeed, in Eq. \ref{ogthm} only the ratio 
of the model dependent parameters in Eq. \ref{maxval}
shows up, thus systematic uncertainties cancel out. In addition, our results 
are not affected by the electron screening, which can enhance the cross section
by a factor $\gtrsim 2.4$ at 20 keV \cite{ASS87}, thus spoiling any direct 
measurement of this resonance. At higher temperatures, higher energy resonances
in the $^{18}{\rm O}(p,\alpha){}^{15}{\rm N}$ reaction can play a role. 
Their study will be the subject of forthcoming works.

\end{document}